# Chaos in Simple Rotation-Translation Models[*]

**Christos H. Skiadas**[1] **and Charilaos Skiadas**[2]

[1]Technical University of Crete, Greece ( Email: skiadas@ermes.tuc.gr )
[2]Hanover College, Indiana, USA ( Email: skiadas@hanover.edu )

**Abstract.** The chaotic properties of simple two-dimensional rotation-translation models are explored and simulated. The models are given in difference equation forms, while the corresponding differential equations systems are studied and the resulting trajectories in the plane are explored and illustrated in the computer experiments done. Characteristic patterns, egg-shaped forms and central chaotic bulges are present when particles are introduced in the rotating system. The resulting forms and chaotic attractors mainly depend on the form of the nonlinear function expressing the rotation angle. Several cases are studied corresponding to a central force rotation system.

## 1 Introduction

Long ago it was proven that any two-dimensional system expressed by a set of two differential equations could not give chaotic paths in a plane. This is the consequence of the Poincaré-Bendixon theorem. However, two difference equations with at least one non-linear term could under certain assumptions give chaotic forms in a plane. In this paper we examine the chaotic forms and attractors, which appear when relatively simple rotation-translation equation forms are applied. Rotation and translation are the key factors for many object formations in nature. The related analysis has been studied mainly from the continuous point of view by means by the differential equations analogue. Here, the approach follows the discrete analogue modeled by a set of difference equations. The notation used starts from the simple parametric form representing a circle

$$x_n = r_n \cos(\theta_n)$$
$$y_n = r_n \sin(\theta_n)$$

where $r_n = \sqrt{x_n^2 + y_n^2}$ and $\theta_n$ is the rotation angle

If the system from step *n* at time *t* is rotated to step *n*+1 at time *t*+1 with an angle $\Delta\theta$ the above equations yield

$$x_{n+1} = r_n \cos(\theta_n + \Delta\theta)$$
$$y_{n+1} = r_n \sin(\theta_n + \Delta\theta)$$

or

---

[*] First version circulated May 5, 2005. Last version January 3, 2007

$$x_{n+1} = r_n[\cos(\theta_n)\sin(\Delta\theta) - \sin(\theta_n)\sin(\Delta\theta)]$$
$$y_{n+1} = r_n[\cos(\theta_n)\sin(\Delta\theta) + \sin(\theta_n)\cos(\Delta\theta)]$$

By using the parametric equations of the circle yields the rotation map in difference equation form

$$x_{n+1} = x_n \cos(\Delta\theta) - y_n \sin(\Delta\theta)$$
$$y_{n+1} = x_n \sin(\Delta\theta) + y_n \cos(\Delta\theta)$$

## 2  Rotation-Translation

When the rotation is followed by a translation equal to $a$, parallel to the $X$-axis, the last equations lead to the following rotation-translation difference equations

$$x_{n+1} = a + x_n \cos(\Delta\theta) - y_n \sin(\Delta\theta)$$
$$y_{n+1} = x_n \sin(\Delta\theta) + y_n \cos(\Delta\theta)$$

The time elapsed between two successive iterations is $\Delta t$ and usually is taken equal to unity. Then by taking into account that

$$\dot{x} = dx/dt \approx \Delta x/\Delta t = x_n - x_{n-1}$$

and

$$\dot{y} = dy/dt \approx \Delta y/\Delta t = y_n - y_{n-1}$$

the resulting differential equations are

$$\dot{x} = a + x[\cos(\Delta\theta) - 1] - y\sin(\Delta\theta)$$
$$\dot{y} = x\sin(\Delta\theta) + y[\cos(\Delta\theta) - 1]$$

## 3  A Simple Rotation-Translation System of Differential Equations

If the angle $\Delta\theta \ll 1$ the last system of differential equations is simplified to the following

$$\dot{x} = a - y\Delta\theta$$
$$\dot{y} = x\Delta\theta$$

From this system the following differential equation yields

$$\frac{dy}{dx} = \frac{x\Delta\theta}{a - y\Delta\theta}$$

4This equation leads to the following form

$$ady = (\Delta\theta)rdr$$

The solution of this differential equation depends on the form of the function that follows $\Delta\theta$. An approach is to consider $\Delta\theta$ as a function of the distance $r$ from the origin. From mechanics we know functions connected with to rotation angle. In the case of a central force [1], [6], from the transverse component of the acceleration, the function $r^2\dot\theta = h_1$ is obtained ($h_1$ is a constant). Provided that $\Delta t = 1$, the following approximation for $\Delta\theta$ is obtained

$$\Delta\theta \approx \frac{h_1}{r^2}$$

Another case appears when a mass rotates in the periphery of a circle under a central force: $f(r) = MG/r^2$. The equation of motion gives

$$\dot\theta = \sqrt{\frac{MG}{r^3}}$$

Thus an approximation for $\Delta\theta$ is ($\Delta t = 1$)

$$\Delta\theta \approx \sqrt{\frac{MG}{r^3}}$$

Finally we proceed to the solution of the following equation

$$ady = \sqrt{\frac{MG}{r}}dr$$

The solution is

$$r = \frac{(ay+h)^2}{4MG}$$

$h$ is an integration constant.

The last equation is transformed in the following form

$$x^2 = \left[\frac{(ay+h)^2}{4MG} - y\right] \cdot \left[\frac{(ay+h)^2}{4MG} + y\right]$$

Exploring the properties of the last equation one finds that $x = 0$, when $y = MG/a^2$ and $h = MG/a$. This value of $h$ applied in the above equation gives an equation for the path of a trajectory in the $(x, y)$ plane, which divides the plane in two segments. This trajectory is the outer limit of the vortex region of the rotation. When $h > MG/a$ the trajectories diverge and the rotating object leads to infinity. When $h < MG/a$ the rotating mass moves inside the space the limits of which are set by the above trajectory.

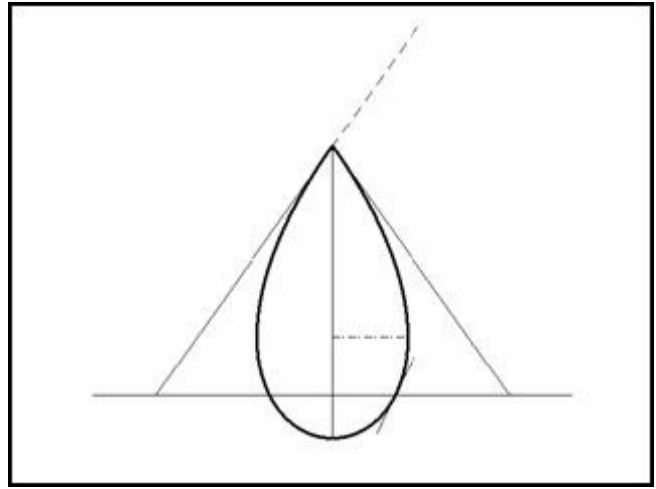

Fig.1 The characteristic trajectory

Some very interesting properties of the two-dimensional function are illustrated in the above figure. The trajectory in the limit of escape has an egg-shape. The sharper corner is at the maximum value of $y = MG/a^2$ where $x = 0$. These values of $x$, $y$ set the derivatives $\dot{x}$, $\dot{y}$ equal to zero. However this point is not stable.

The minimum value of $y$ is that where the maximum rotation speed is achieved, $y = MG/a^2 (\sqrt{2} - 1)^2$, $x = 0$.

The maximum

$$x_{max} = \frac{MG}{a^2} \sqrt{\frac{5\sqrt{5} - 11}{2}} \approx 0.3 \frac{MG}{a^2}$$

is estimated by equating to zero the first derivative of $x$ with respect to $y$ that is

$$\dot{x} = \frac{1}{x}\left[\frac{a}{8(MG)^2}\left(ay + \frac{MG}{a}\right)^3 - y\right]$$

This is achieved when

$$y = \frac{MG}{a^2}(\sqrt{5} - 2)$$

When $y = 0$ then $x = MG/(4a^2)$. The tangent at this point is estimated by setting $y = 0$ to the first above derivative of $x$ with respect to $y$ that is $dx/dy = 1/2$ or $dy/dx = 2$. This tangent appears in the above figure.

The tangent in the top sharp corner of the egg-shape form is more difficult to be estimated because the value of the first derivative of $x$ with respect to $y$ above is of the order 0/0 and so it is for the second derivative. Thus, a small change $\varepsilon$ is added to $y$ so that the value of this variable is $y = c/a^2 + \varepsilon$. In the following $x$ is estimated from the original equation for $(x, y)$ provided that the higher values of $\varepsilon$ are dropped. The result is $x = \varepsilon\sqrt{2}/2$. The tangent at the top point is $dy/dt \approx \Delta y/\Delta x = \varepsilon/(\varepsilon\sqrt{2}/2) = \sqrt{2}$.

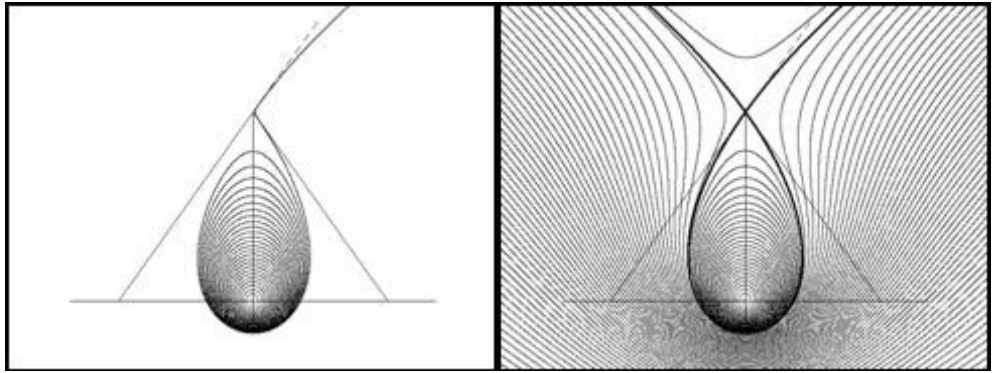

Fig.2 Rotating paths          Fig.3 Paths in the plane

The figure to the left illustrates the path of a moving particle. The particle follows an anti-clockwise direction. It starts from a point near zero and rotates away from the center until the highest permitted place. Then, it escapes to infinity following the tangent in top of the egg-shaped path. In the right figure a large number of paths are drown. Every path has the same value for $h$. The paths outside the egg-shaped limits have direction from left to the right and they do not transverse the egg-shaped region of the plane. In this case the value of $h = MG/a^2$. The iterative formula for the above simulations is the following

$$x_{n+1} = x_n + \left(a - y_n\sqrt{\frac{MG}{r_n^3}}\right)d$$

$$y_{n+1} = y_n + x_n\sqrt{\frac{MG}{r_n^3}}d$$

$$r_n = \sqrt{x_n^2 + y_n^2}$$

The parameters in both figures are $a = 2$, $MG = 100$ and $d = 0.05$.

Other interesting properties of the above egg-shaped forms arise by using elements from mechanics. By observing the equation of motion it is clear that the parameter $a$ expresses a constant speed with direction parallel to the $X$-axis. The rotational speed has two components, $\dot{x}, \dot{y}$. However at the top and at the bottom of the egg-shaped form only the $x$ component of the velocity is present. The rotational speed or the transverse component of the velocity is $\upsilon = \sqrt{MG/y}$. This speed must be equal to $a$. Thus, without using the equations for the trajectories the value of $y$ at the top, $x = 0$, is estimated from the equality

$$\upsilon - a = 0$$

Thus

$$\sqrt{\frac{MG}{y}} = a$$

and

$$y = \frac{MG}{a^2}$$

To obtain the equation for the total speed at the lower point of the curve bottom we must take into account that this speed must be equal to the escape speed

$$\upsilon + a = \upsilon_{esc}$$

The speed of escape is

$$\upsilon_{esc} = \sqrt{\frac{2MG}{y}}$$

Thus, the equation becomes

$$\sqrt{\frac{MG}{y}} + a = \sqrt{\frac{2MG}{y}}$$

Finally from this last equation the value of $y$ results

$$y = \frac{MG}{a^2}(\sqrt{2}-1)^2$$

This is precisely the same result to that by using the equation obtained earlier for the trajectories of the rotation system. It is obvious that the system in these special cases obeys the lows we learn from classical mechanics. This egg-like form is a set of vortex curves with relative stability and strength.

## 4  A Discrete Analog of the Rotation-Translation Model

The analysis above was important in order to be able to understand the dynamics underlying the rotation analog in the discrete case. The iterative scheme based on the difference equations for rotation-translation is more difficult to handle analytically but, from the other hand, is more close to the real situation. No one approximation is needed, whereas, the egg-shaped scheme and the trajectories are retained. Moreover, the trajectories inside the egg-shaped boundary are perfectly closed loops. Also, close to the center, the chaotic nature of the rotation-translation procedure is presented. In the original equations presented earlier we introduce an area contracting parameter $b$ $(0 < b < 1)$ and the final equations used are the following

$$x_{n+1} = a + b[x_n \cos(\Delta\theta) - y_n \sin(\Delta\theta)]$$
$$y_{n+1} = b[x_n \sin(\Delta\theta) + y_n \cos(\Delta\theta)]$$

where, the same approximation for

$$\Delta\theta_n \approx \sqrt{\frac{MG}{r_n^3}}$$

is accepted.

A convenient formulation of the last iterative map expressing rotation and translation is by using the following difference equation

$$f_{n+1} = a + bf_n e^{i\Delta\theta}$$

where

$$f_n = x_n + iy_n$$

The determinant of this map is det $J = b^2 = \lambda_1\lambda_2$. This is an area contracting map if the Jacobian $J$ is less than 1. The eigenvalues of the Jacobian are $\lambda_1$ and $\lambda_2$, and the translation parameter is $a$.
In the case selected above rotation is followed by a translation along the $X$ axis. Applying a translation only to one axis instead to both $X$ and $Y$ has the advantage of dropping one parameter while this do not affect the results in the majority of the cases studied.
A very interesting property of the map is the existence of a disk $F$ of radius $ab/(1-b)$ centered at $(a, 0)$ in the $(x, y)$ plane. The points inside the disk remain inside under the map.
The fixed points obey the following equations

$$\left(x - \frac{a}{1-b^2}\right)^2 + y^2 = \frac{a^2 b^2}{(1-b^2)^2}$$

and

$$\cos(\Delta\theta) = \frac{1}{2b}\left(1 + b^2 - \frac{a^2}{r^2}\right)$$

The first equation defines a circle K with radius $R = ab/(1-b^2)$, centered at $[a/(1 - b^2), 0]$ in the $(x, y)$ plane. The second equation can be solved provided that the angle $\Delta\theta$ is a function of $r$.

The first part of the analysis is based on the assumption that no-area contraction takes place so that the parameter $b = 1$. In this case the angle of the disk $F$ tends to infinity and thus, all the space of the plane $(x, y)$ is considered. An immediate finding from the analysis is that the symmetry axis of the map is that expressed by the line $x = a / 2$. This is different from the symmetry axis of the differential equation analogue where the symmetry axis is the line $x=0$. In the difference equation case the egg-shaped form is moved to the right from the original positions of the system of coordinates. The simplest way to find the axis of symmetry is to search for the points where $x_{n+1} = x_n$ and $y_{n+1} = y_n$. This leads to the following

$$x = \frac{a}{2}$$

and

$$y = \frac{a}{2}\cot\left(\frac{\Delta\theta}{2}\right)$$

It is clear that there is no-symmetry axis in the *X* direction as *y* does not show stable magnitude, as it is a function of the angle $\Delta\theta$. However an approximation of *y* is achieved when the angle $\Delta\theta$ is small. Then

$$y \approx \frac{a}{\Delta\theta} = \frac{MG}{a^2}.$$

This point ($x = a/2$, $y = MG/a^2$) is an equilibrium point. However, this is not a stable equilibrium point. The system easily may escape to infinity when *y* is higher from this last value [9].

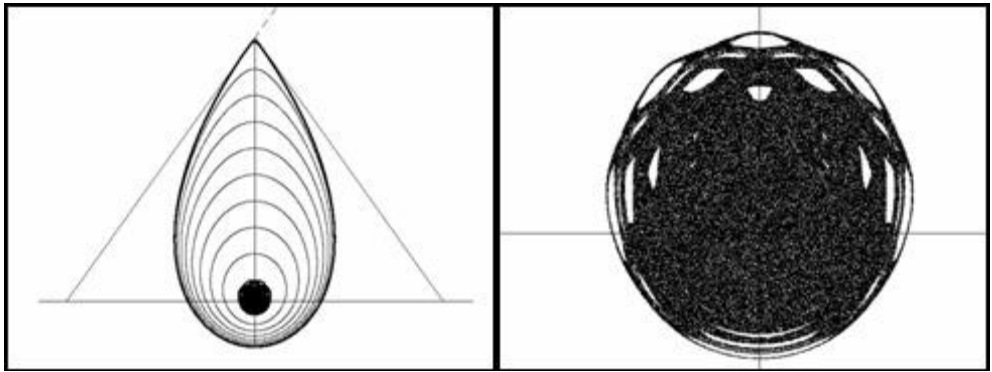

Fig.4 Rotation-translation          Fig.5 The central chaotic bulge

The above figures illustrate a rotation-translation case with parameters $a = 2$, $MG = 100$. In the figure to the left a number of paths inside the egg-shaped formation are drown. All the paths are analogous to those obtained by the differential equation analogue studied above but they are moved to the right at a distance $x = a/2$. A chaotic attractor like a bulge appears around the center of coordinates. An enlargement of this attractor is illustrated in the figure to the right. This is a symmetric chaotic formation with axis of symmetry the line $x = a/2$.

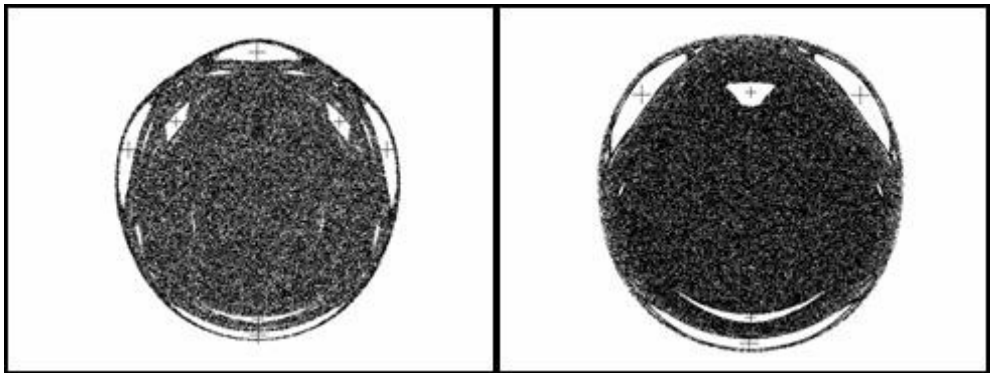

Fig.6 The Chaotic bulge (*a* = 1)          Fig.7 The chaotic bulge (*a* = 0.8)

Illustration of the chaotic bulge of the rotation-translation model is illustrated in the above Figures. The parameters selected are, *a* = 1 for the left figure and *a* = 0.8 for the right figure and *MG* = 100 for both cases. The rotation angle is $\Delta\theta = \sqrt{MG/r^3}$.

In the left figure third (small cross) and fourth order equilibrium points appear indicated by a cross. In this case the following relations hold: ($x_{n+4} = x_n$, $y_{n+3} = y_n$) and ($x_{n+4} = x_n$, $y_{n+4} = y_n$). Accordingly in the right figure two equilibrium cases appear, one of second order (indicated by small cross) where $x_{n+2} = x_n$ and $y_{n+2} = y_n$ and one of the third order where $x_{n+3} = x_n$ and $y_{n+3} = y_n$.

In both cases an algorithm is introduced for the estimation of the equilibrium points. The convergence is quite good provided that the starting values are in the vicinity of the equilibrium points.

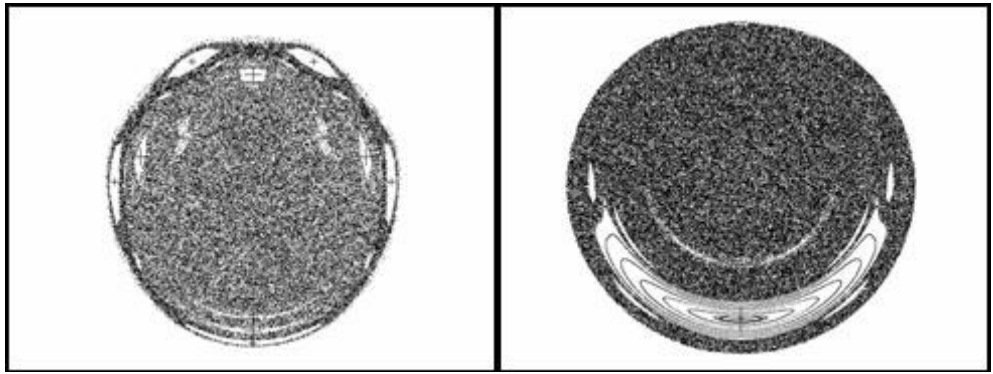

Fig.8 The chaotic bulge ($a = 1.2$)    Fig.9 The chaotic bulge ($a = 0.2$)

As the parameter $a$ is changed to higher values higher order equilibrium points appear. A fifth order equilibrium point is added to those presented before as is illustrated in the above left figure where $a = 1.2$. The relations for the new equilibrium point are $x_{n+5} = x_n$ and $y_{n+5} = y_n$. The figure to the right illustrates the case when $a = 0.2$. Only a first order equilibrium point is present, $x_{n+1} = x_n$ and $y_{n+1} = y_n$.

Higher order bifurcation is present in the next two figures. In the left, a tenth order bifurcation form appears in the periphery of the chaotic attractor. The magnitude of the translation parameter is $a = 1.5$. In the right figure the translation parameter is $a = 1.9$. At this value the bifurcation is so high that the chaotic region covers all the space of the plane that is included from the periphery of the egg-shaped form. The point mass, following chaotic trajectories finally escape to infinity.

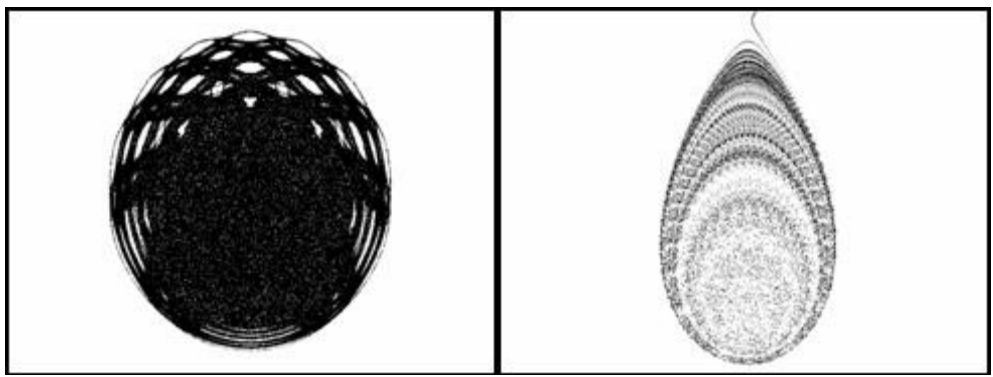

Fig.10 The chaotic bulge ($a = 1.5$)    Fig.11 Complete chaos ($a = 1.9$)

A difficult problem is to find the coordinates of equilibrium-points, which are located in the center of the islands that are in the middle of the chaotic sea of the attractor.
First, it is easy to verify that an equilibrium point must be located on the axis of symmetry if it is a sole point or if it is included in a group of points and the number of these points is odd.

In this case all the other points are divided in symmetric pairs and are located in positions outside of the symmetry axis $(x = a/2)$.

A first order equilibrium point is characterized by the simple relations $x_{n+1} = x_n$ and $y_{n+1} = y_n$ which lead to the following relation for the radius of the circle on the periphery of which the equilibrium point must be located

$$R_k \approx \sqrt[3]{\frac{MG}{(2k\pi)^2}}, \quad k = 1, 2, \ldots$$

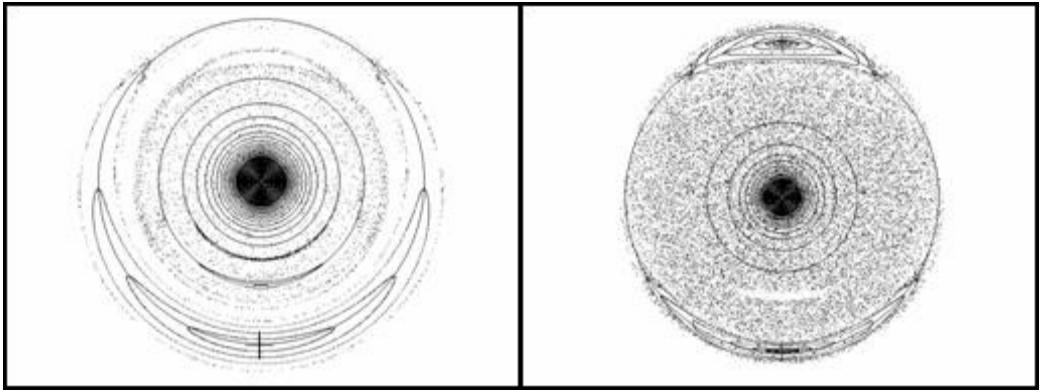

Fig.12 Equilibrium points ($a = 0.1$)   Fig.13 Equilibrium points ($a = 0.5$)

The circles $R_k$ introduce a set of decreasing order as regards the magnitude of the radius, as is illustrated in the left figure above ($a = 0.1$). The first point located on the circle with the larger radius ($k = 1$) is indicated by a cross. Around this points are designed few isoclines indicating the limits of the island. The next and the following points follow in decreasing order. They are located in the middle of the isoclines.

The equilibrium points of higher order follow by using different relations. A simple case is that of the estimation of the equilibrium points of the second order that is when the coordinates are, $x_{n+2} = x_n$ and $y_{n+2} = y_n$. This assumption leads to the following equation

$$(x^2 + y^2)\cos^2(\Delta\theta) + 2x^2 \cos(\Delta\theta) + x^2 - y^2 = 0$$

This equation is fulfilled if the rotation angle obeys the relation $\Delta\theta = (2k+1)\pi$. Then this result leads to the following relation for the radius of the circle on the periphery of which the equilibrium points must be located.

$$R_k \approx \sqrt[3]{\frac{MG}{((2k+1)\pi)^2}}, \quad k = 1, 2, \ldots$$

The simulation results are illustrated in the right figure above. The translation parameter is $a=0.5$. The co-centric circles are distributed in decreasing order of the magnitude of the radius $R_k$. The first two equilibrium points are distributed in the outer circle. They are indicated by a cross and are located on the axis of symmetry, $x = a / 2$, but in opposite directions one to the positive and one to the negative part of the semi-plane for $y$. Three pairs of order two equilibrium points are illustrated in the next left figure. The equilibrium pairs are indicated by a cross. Also the three equilibrium islands of the first order are presented. The parameter $x=0.04$. As this parameter has a very low value, the central chaotic bulge has a small radius.

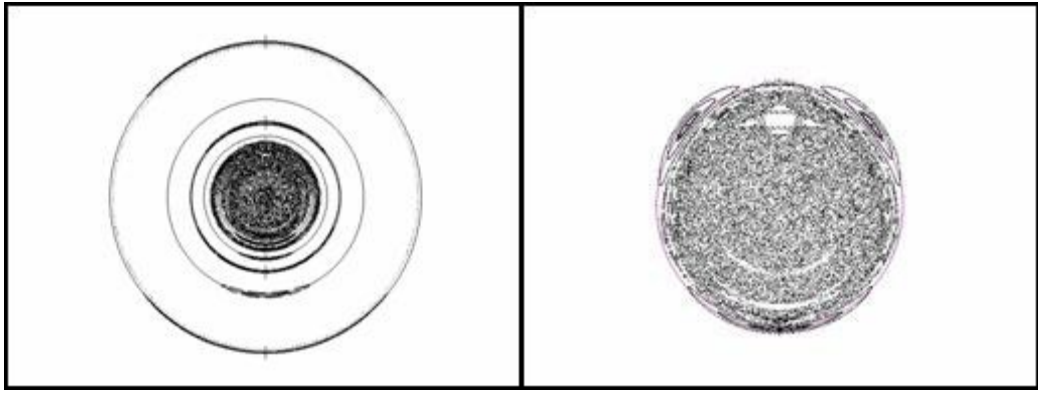

Fig.14 Equilibrium pairs (*a* = 0.04)  Fig.15 Order-3 bifurcation (*a* = 0.7)

The order three equilibrium points are presented in the right figure. They are located in the middle of the three islands, which are in the outer part of the chaotic bulge. The translation parameter is *a* = 0.7. The order three equilibrium points are in the periphery of a circle centered at (*a*/2, *a*/5). The radius of this circle is approximated by

$$R \approx \sqrt[3]{\frac{MG}{(2\pi/3)^2}} - \left(\frac{a}{4}\right)^2$$

## 5  A General Approach of the Rotation-Translation Model

The solution of the differential equation of this model leads to the form

$$ay + h = \int (\Delta\theta) r dr$$

A simple approximation of the quantity $\Delta\theta$ is the following

$$\Delta\theta \approx \sqrt{\frac{GM}{r^\beta}}$$

Then the solution has the form

$$ay + h = \frac{r^{2-\beta/2}}{2-\beta/2}\sqrt{GM}$$

The special case $\beta = 2$ leads to the following equation form

$$ay + h = r\sqrt{GM}$$

or

$$(x^2 + y^2)GM = (ay + h)^2$$

This equation expresses a family of ellipses. This family is illustrated in the next figure to the left. The parameter $a = 6$ and the constant of integration $h$ takes several values. The axis of symmetry is at $x = a / 2$. The figure to the right illustrates a chaotic central bulge. The outer part of this chaotic attractor is approximated by an ellipse with $h = MG + a^2$.

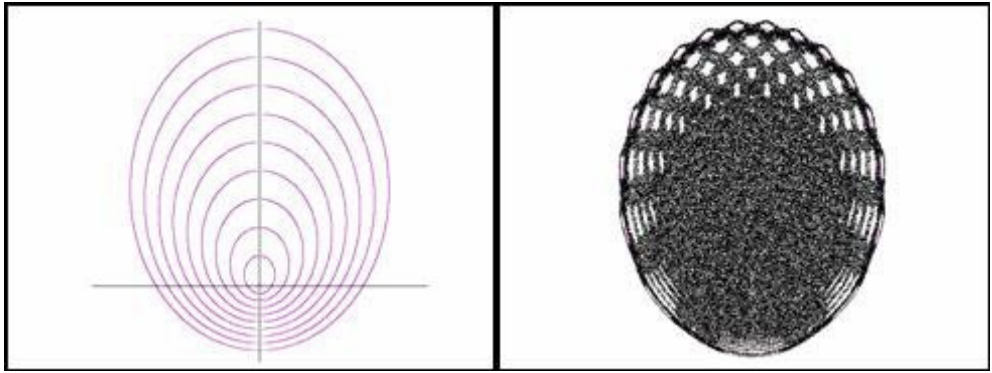

Fig.16 A family of ellipses          Fig.17 The chaotic bulge

When $\beta = 4$ the resulting equation for $(x, y)$ is

$$\sqrt{MG}\ln(r) = ay + h$$

The rotating forms are egg-shaped and similar to those provided in earlier rotation forms. With the exception of $\beta = 2$, when the rotation paths are elliptic, the cases with $\beta > 2$ provide egg-shaped rotation forms.

Another very important property of the elliptic case ($\beta = 2$) is that the rotating particles remain in the elliptic paths even if high values for the parameter $h$ are selected. In the limit when the parameter $a$ approaches zero the ellipses turn to be co-centric circles.

## 6 Rotating Particles inside the Egg-shaped Form

Two cases are of particular importance. In the first case the translation parameter $a$ is quite small and the particles are trapped inside the egg-shaped form and remain there following trajectories proposed by the theory presented above. However, a region inside the trapping region is characterized as the chaotic region. In this region the particles follow chaotic paths that form the attractor forms presented above. In the next figure to the left the outer limits of the egg-shaped form are drawn and a disk of rotating particles of equal mass is centered at (0,0). The particles are distributed by following the inverse law given by $\rho = c_1 / r^3$. The diameter of the disk is chosen as to be exactly within the limits of the egg-shaped form. The parameters are $a = 0.25$ and $GM_0 = 0.45$. The rotation angle is given by

$$\Delta\theta = \sqrt{\frac{GM_0}{r^3}}$$

The resulting form that the disc of rotating particles takes after time $t = 10$ appears in the right figure. The original cyclic form is now changed providing an outer form of rotation and an inner chaotic attractor-like object.

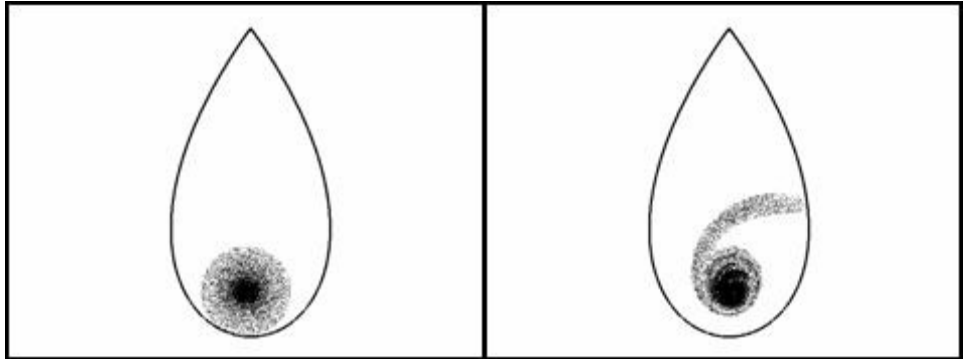

Fig.18 Rotating particles ($t = 0$)    Fig.19 Rotating particles ($t = 10$)

The next figures bellow (left and right) illustrate the resulting picture after time $t = 20$ and $t = 100$ respectively. The distinct inner attractor is more clearly formulated. In the case presented in the right figure the outer part of the rotating object is split in almost all the space of the egg-shaped form but by following characteristic paths.

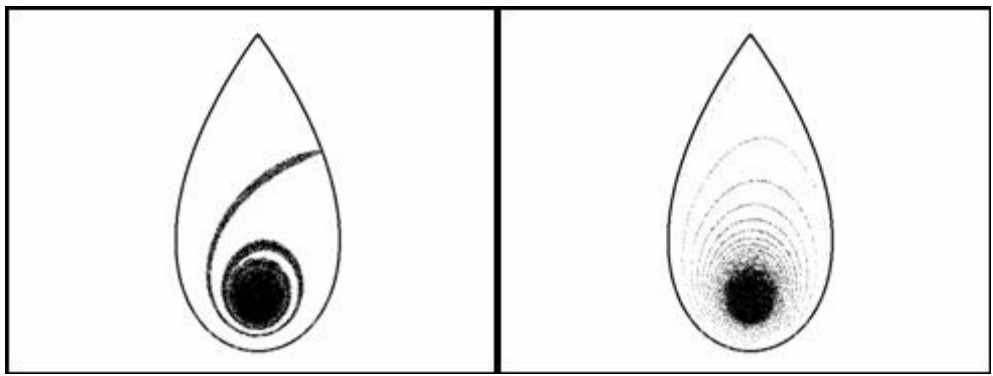

Fig.20 Rotating particles ($t = 20$)    Fig.21 Rotating particles ($t = 100$)

In the second case the translation parameter is high enough as to all of the egg-shaped space is included in the chaotic region. The system is unstable and the rotating particles are not retained inside the egg-shaped region. Instead, they escape by following the escape trajectories and they move away of the egg-shaped region. After some time the majority of the particles will live the region. In the following, the first three figures illustrate, from left to right, three instances after time $t = 2$, $t = 3$ and $t = 4$ respectively. The translation parameter is $a = 0.6$. The cross in the figure to the left is in the position $(x = a, y = 0)$. A cross indicates the characteristic centers of chaotic forms. These coordinates are calculated by using as starting values in a repeated procedure $x = a$ and $y = 0$. The repeated procedure is based on the difference equations for $x$ and $y$.

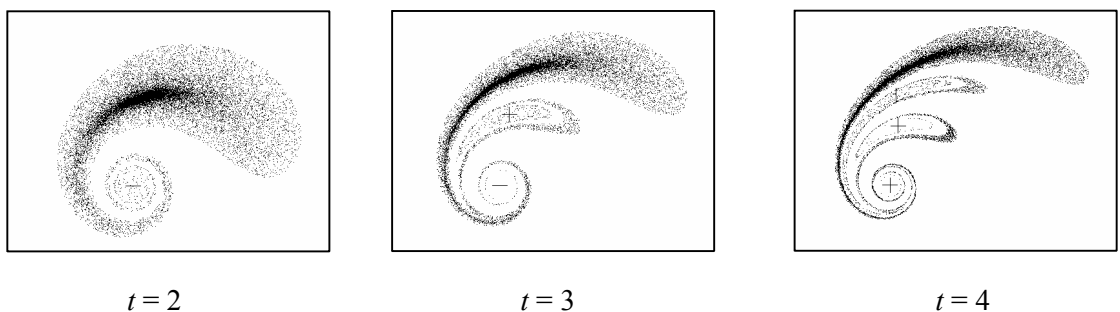

$t = 2$           $t = 3$           $t = 4$

Fig.22 Development of rotating particles during time. The translation parameter is $a = 0.6$.

The case when *t* = 10 appears in the next figure. The rotating system of particles is in an intermediate stage. It escapes from the egg-shaped form but a part of the particles remains around the location (*x* = *a*, *y* = 0) inside the egg-shaped form. When the magnitude of the translation parameter takes higher values, all the particles escape after a short time interval. An intermediate stage appears in the middle figure. The parameter *a* = 0.7 and the time *t* = 10. The leave like structure starts to disappear inside the egg-shaped form and close to the location (*x* = *a*, *y* = 0). These jets eject material through the trajectories of the model. The speed of these jets depends on the magnitude of the parameter *a*, and on the original rotation speed. The speed and the direction of the jets of material coincides to *a* after a large enough amount of time *t*. The next figure to the right illustrates the case when *a* = 0.8. Only the front part of the jet remains connected to the source of the chaotic sea inside the egg-shaped pattern.

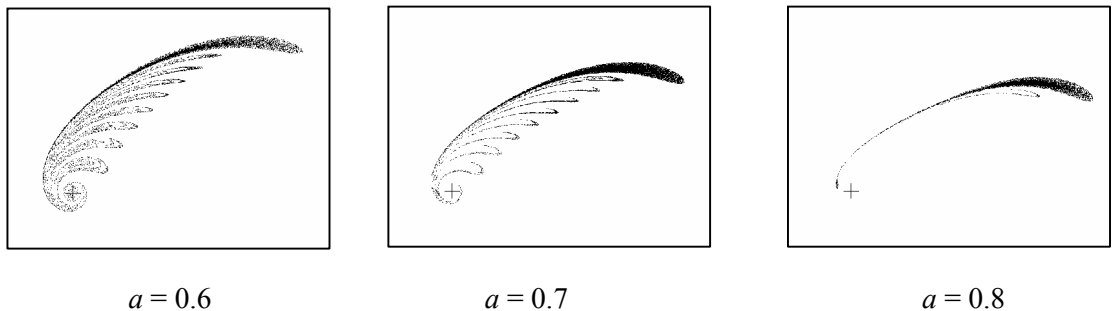

*a* = 0.6          *a* = 0.7          *a* = 0.8

Fig.23 Development of rotating particles according to the magnitude of the translation parameter. The elapsed time is *t* = 10.

When the translation parameter *a* = 1 (*t* = 10) all the rotating material escapes outside the egg-shaped pattern. In the figure presented below (left) the original cyclic disk was quite large and a part of this disk lied outside of the egg-shaped formation. However all the material follows the escape route as a compact formation. Instead in the other figure (right) the influence of a small translation parameter (*a* = 0.25) to large original cyclic cloud has a critical impact to the cloud formation. After *t* = 100 the cloud is separated to two formations, the first inside the egg-shaped form with the chaotic bulge-form in the middle and the second outside.

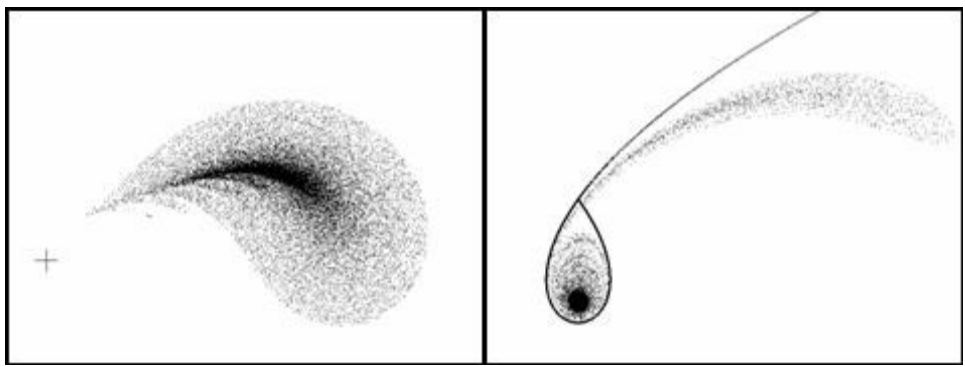

Fig.24 Rotations (*a* = 1, *t* = 10)      Fig.25 Rotations (*a* = 0.25, *t* = 100)

## 7  Rotations Following an Inverse Square Law

We return to the original system of differential equations and the resulting differential equation for *x, y*

$$\dot{x} = a - y\Delta\theta$$
$$\dot{y} = x\Delta\theta$$

and

$$\frac{dy}{dx} = \frac{x\Delta\theta}{a - y\Delta\theta}$$

This equation leads to the following form

$$ady = (\Delta\theta)rdr$$

The solution of this differential equation depends on the form of the function that follows the rotation angle $\Delta\theta$. In this approach we use a function known from mechanics dealing with the rotation angle. This function arises from the law related to the transverse component of the acceleration in a circular movement. The function is expressed by the formula $r^2\dot{\theta} = c_1$ ($c_1$ is a constant). Provided that $\Delta t = 1$, the following approximation for $\Delta\theta$ is obtained

$$\Delta\theta \approx \frac{c_1}{r^2}$$

Finally we proceed to the solution of the following equation

$$ady = \frac{c_1}{r^2}dr$$

The solution is

$$c_1 \ln r = ay + h$$

$h$ is an integration constant.

The last equation is transformed in the following form

$$\frac{c_1}{2}\ln(x^2 + y^2) = ay + h$$

Exploring the properties of the last equation we set $x = 0$ when $y = r$. In this case the maximum or minimum values of $y$ are obtained. The resulting equation for $y$ is

$$c_1 \ln y = ay + h$$

The maximum value for $y$ is achieved when $y = c_1/a$. Then, the value of the parameter $h$ at this limit is

$$h_{crit} = c_1\left[\ln\frac{c_1}{a} - 1\right]$$

This value of $h_{crit}$ applied in the above equation gives an equation for the path of a trajectory in the $(x, y)$ plane, which divides the plane in two segments. This trajectory is the outer limit of the vortex region of the rotation. When $h > h_{crit}$ the trajectories diverge and the rotating object escapes to infinity. When $h < h_{crit}$ the rotating mass rotates inside the limits set by the above trajectory.

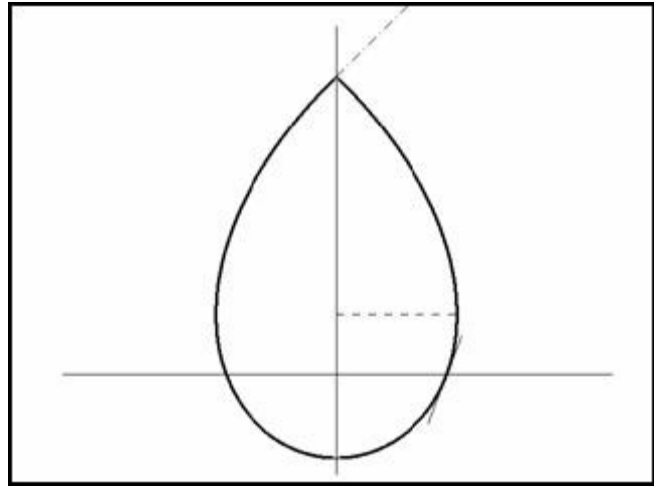

Fig.26 The characteristic trajectory

Some very interesting properties of the two-dimensional function are illustrated in the above figure. The trajectory in the limit of escape has an egg-shape. The sharper corner is at the maximum value of $y = c_1/a$ where $x = 0$. These values of $x, y$ set the derivatives $\dot{x}, \dot{y}$ equal to zero. However this point is not stable. This maximum value is obtained by solving the following equation for $y$

$$\ln\left(\frac{ay}{c_1}\right) - \frac{ay}{c_1} + 1 = 0$$

The minimum value of $y$ is that where the maximum rotation speed is achieved. The resulting equation for $y$ is

$$\ln\left(\frac{ay}{c_1}\right) + \frac{ay}{c_1} + 1 = 0$$

A numerical solution gives $y = 0.278 c_1/a$.

The maximum $x_{max}$ is estimated by equating to zero the first derivative of the following equation for $x, y$

$$\frac{c_1}{2}\ln(x^2 + y^2) = ay + c_1 \ln\left(\frac{c_1}{a}\right) - c_1$$

After appropriate differentiation, a numerical solution of the resulting equation gives: $x_{max} = 0.402 c_1/a$. This is achieved at $y = 0.203 c_1/a$.

When $y = 0$ then $x = c_1/(ae)$. The tangent at this point is $dy/dx = e$. This tangent appears in the above figure.

The tangent in the top sharp corner of the egg-shaped form is more difficult to be estimated because the value of the first derivative of *x* with respect to *y* above is of the order $0/0$ and so it is for the second derivative. Thus, a small change $\varepsilon$ is added to *y* so that the value of this variable is $y = c_1/a + \varepsilon$. In the following *x* is estimated from the original equation for (*x*, *y*) provided that the higher values of $\varepsilon$ are dropped. The result is $x = \varepsilon$. Then the tangent at the top point is $dy/dx = 1$.

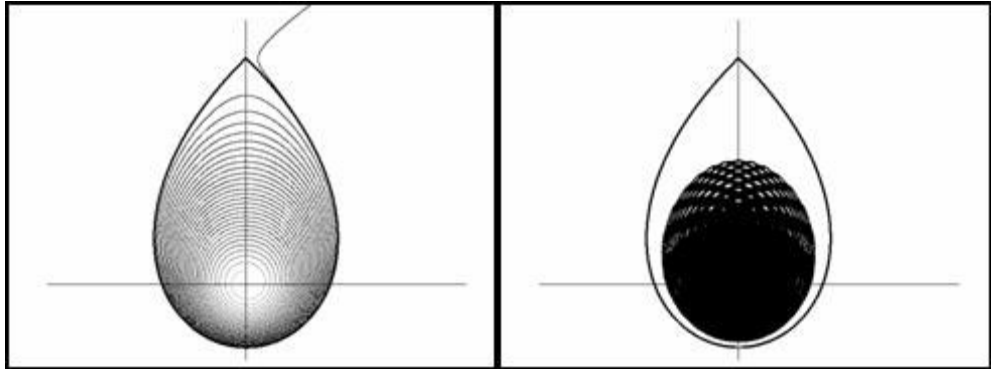

Fig.27 The continuous case    Fig.28 The central chaotic bulge

In the two figures above important properties of the model are presented. In the left figure the continuous model presented below is simulated. Instead, in the figure to the right, the discrete analogue of the model appears. The figure to the right illustrates the central bulge in the case of $a = 2.7$ and $c_1 = 100$. The value of the translation parameter is quite high. This causes the appearance of an enormous chaotic attractor in the middle of the egg-shaped area. As in the case presented earlier strong bifurcation is present.

As is illustrated in the above figure (left) the particle follows an anti-clockwise direction. It starts from a point near zero and rotates away from the center until the highest permitted place. Then, it escapes to infinity following the tangent in top of the egg-shaped path. The iterative formula for the above simulations is the following

$$x_{n+1} = x_n + \left(a - y_n \frac{c_1}{r^2}\right)d$$

$$y_{n+1} = y_n + x_n \frac{c_1}{r^2} d$$

$$r_n = \sqrt{x_n^2 + y_n^2}$$

The parameters in both figures are $a = 2.7$, $c_1 = 100$ and $d = 0.0001$. A Runge-Kutta method of fourth order could also apply.

# 8 Conclusions

Several cases involving Rotation and Translation are investigated and simulated in this paper. A large variety of special problems, starting from rotations under a central force, fluid dynamics and vortex flow ([2], [3], [5], [10]) to formations in astronomy and astrophysics [4], [7], [8], can be simulated by using very simple chaotic and non-chaotic models as those indicated in the chaotic experiments done. Characteristic patterns, egg-shaped forms and central chaotic bulges are present when particles are introduced in the rotating system. The resulting forms and chaotic attractors mainly depend on the form of the nonlinear function expressing the rotation angle. The presented material is a part of a larger study involving the chaotic modeling and simulation field. Further results will be given in other communications.